\documentclass[
onecolumn,
]{ceurart}

\usepackage{xcolor}
\usepackage{url}
\usepackage{subfigure}

\begin{document}

\copyrightyear{2022}
\copyrightclause{Copyright for this paper by its authors.
  Use permitted under Creative Commons License Attribution 4.0
  International (CC BY 4.0).}

\conference{SEBD 2022: The 30th Italian Symposium on Advanced Database Systems, June 19-22, 2022, Tirrenia (PI),
Italy
}

\title{Extracting Large Scale Spatio-Temporal Descriptions from Social Media}

\author{Carlo Bono}[
orcid=0000-0002-5734-1274,
email=carlo.bono@polimi.it,
]
\author{Barbara Pernici}[
orcid=0000-0002-2034-9774,
email=barbara.pernici@polimi.it,
]

\address{Politecnico di Milano, DEIB,
  Via Giuseppe Ponzio, 34, 20133 Milano, Italia}

\begin{abstract}
The ability to track large-scale events as they happen is essential for understanding them and coordinating reactions in an appropriate and timely manner. This is true, for example, in emergency management and decision-making support, where the constraints on both quality and latency of the extracted information can be stringent. In some contexts, real-time and large-scale sensor data and forecasts may be available. We are exploring the hypothesis that this kind of data can be augmented with the ingestion of semi-structured data sources, like social media. Social media can diffuse valuable knowledge, such as direct witness or expert opinions, while their noisy nature makes them not trivial to manage. This knowledge can be used to complement and confirm other spatio-temporal descriptions of events, highlighting previously unseen or undervalued aspects. The critical aspects of this investigation, such as event sensing, multilingualism, selection of visual evidence, and geolocation, are currently being studied as a foundation for a unified spatio-temporal representation of multi-modal descriptions. The paper presents, together with an introduction on the topics, the work done so far on this line of research, also presenting case studies relevant to the posed challenges, focusing on emergencies caused by natural disasters.
\end{abstract}


\maketitle

\section{Introduction}

The nature of social networks implies an intrinsic distributional variability, since the topics on social networks usually reflect, to different extents, events taking place in the 
world. Some items transport relevant information about some aspect of reality, usually in the form of unstructured and semi-structured data. For example, information about a specific event could be made available using text in some language and/or images and videos, and further contextualized with tags and conversations, extended with external links, and enriched with comments and interactions. These different dimensions of social media posts contain information, explicitly or implicitly, possibly related to actual events. ``Possibly'' since separating relevant\footnote{Relevant as in ``informative with respect to to some chosen event, kind of event or aspect of reality''} and irrelevant information is usually cumbersome, both because of input data quality and volume. Since we aim at extracting descriptions of large-scale events along the dimensions of space and time, we mainly focus on two operational aspects: the ability to automatically detect whether a certain event is actually ongoing, and the ability to extract quantitative, actionable measures as the event unfolds, possibly in an adaptive fashion. The former is meant to be used as a trigger for preempting data collection tasks. The latter has to be compared and complemented with other layers, such as sensor or forecast data, in order to enhance the understanding of reality. Both aspects are affected by the variability of the social media data distribution over space and time, and both can exploit the same variability. 

The remainder of this work is structured as follows. In the next session, the main challenges and topics in the literature are reviewed. Section \ref{sec:issuesmethods} illustrates the investigations conducted so far on these topics. Section \ref{sec:casestudies} combines these results in an application customized for flood events. Finally, Section \ref{sec:futurework} sketches future medium-term research directions.

\section{Related work} \label{sota}

The opportunities and challenges for using  social media in emergency management have been trending topics over the last years. Large-scale emergencies are not trivial to tackle with automated and practical tools. We focus on emergencies since there is an essential value, as well as they pose complex questions to be studied. The papers \cite{IMRAN2020102261, lorinisocial} highlight recent opportunities, challenges, applications and directions for the use of social media data for emergency management at large. Vertical studies for the use of social media for specific events have been performed, such as \cite{DBLP:journals/tkde/SakakiOM13} for earthquake events and \cite{fohringer2015social, shoyama21} for the case of flood events. Techniques for automatic image filtering on social network data, mainly using deep learning, have been studied in the field of emergency and social sensing in \cite{nguyen2017automatic, DBLP:conf/icse/NegriSARSSFCP21}.

Methodologies and resources to analyze social media posts in multiple languages are also essential, as discussed in \cite{Olteanu_Castillo_Diaz_Vieweg_2014} for lexicon building, and in \cite{Autelitano2019} for an adaptive, cross-source approach to social media crawling. How and to which extent social media can be integrated into production-grade emergency management systems has been studied in \cite{stollberg2012use} and \cite{lorini2020social}. Some of the current work builds on past experience with emergencies, as documented in \cite{havas2017e2mc}. 

Although many studies explicitly address detection, filtering, geolocation and multilinguality support, some limitations exist. Production-level requirements are hardly met for many kind of events. Multi-modal approaches are sometimes investigated in literature, but not widely adopted and usually limited to one of the tasks. The same holds true for data fusion approaches. Our investigation aims at designing an end-to-end framework for automatically deriving event characterizations, taking advantage of multi-modal and multi-source approaches in order to solve these challenges. Benchmarking and comparison with existing solutions will be performed on publicly available datasets and self-produced ones.

\section{Issues and methods}
\label{sec:issuesmethods}

\subsection{Event detection and adaptation}\label{subsec:detection}
Monitoring regions of the world for classes of events is one of the possible entry points for social media data collection and processing. Some notable sources of information can dispatch alerts and forecasts, as it happens with GDACS\footnote{The Global Disaster Alert and Coordination System, started in 2004, is a joint initiative of the United Nations Office for the Coordination of Humanitarian Affairs (UN OCHA) and the European Commission (EC).} and GloFAS\footnote{The Global Flood Awareness System is a component of the Copernicus Emergency Management Service (CEMS) that delivers global hydrological forecasts to registered users.}.

In this context, we study both timeliness and geographical precision of information derived from social media, in particular tweets, as we analyze it compared to information available in authoritative sources.
Not all events are covered with the same timeliness, so for large scale events, we are using posterior validated data to understand to which extent social media can support already available data, such as forecast and sensor data, as a detection mechanism.  

As a first step, the use of dictionaries of query terms for social sensing is being evaluated. For each of the dictionary terms, the time series of the term usage over a recent time window can be obtained. This signal is then used to estimate if an event of interest is ongoing or not. A first exploration of this methodology can be found in \cite{bono2022triggercit}, in which we focused on searching Twitter in many possible languages starting from a limited set of seed search keywords. We are studying how to leverage available ground truth in order to automatically build language-tailored resources. The current focus is on building language-centered dictionaries with a data-centered approach\footnote{Leveraging language models, word embeddings and external data sources}, minimizing human intervention and taking advantage of the temporal correlation between words and events. Feature windows built with these dictionaries are then fed to a supervised classifier, such as a CNN, and evaluated in a leave-one-event-out fashion. Our preliminary assessment shows that a low precision (<50\%) and a good recall (\textasciitilde 80\%) triggering system can be achieved, with almost no human intervention. This result is achieved with negligible processing effort, and a comfortable request resolution.\footnote{One request per second \url{https://developer.twitter.com/en/docs/twitter-api/rate-limits}}

\begin{figure}[h!]
\centering
  \includegraphics[width=0.65\columnwidth]{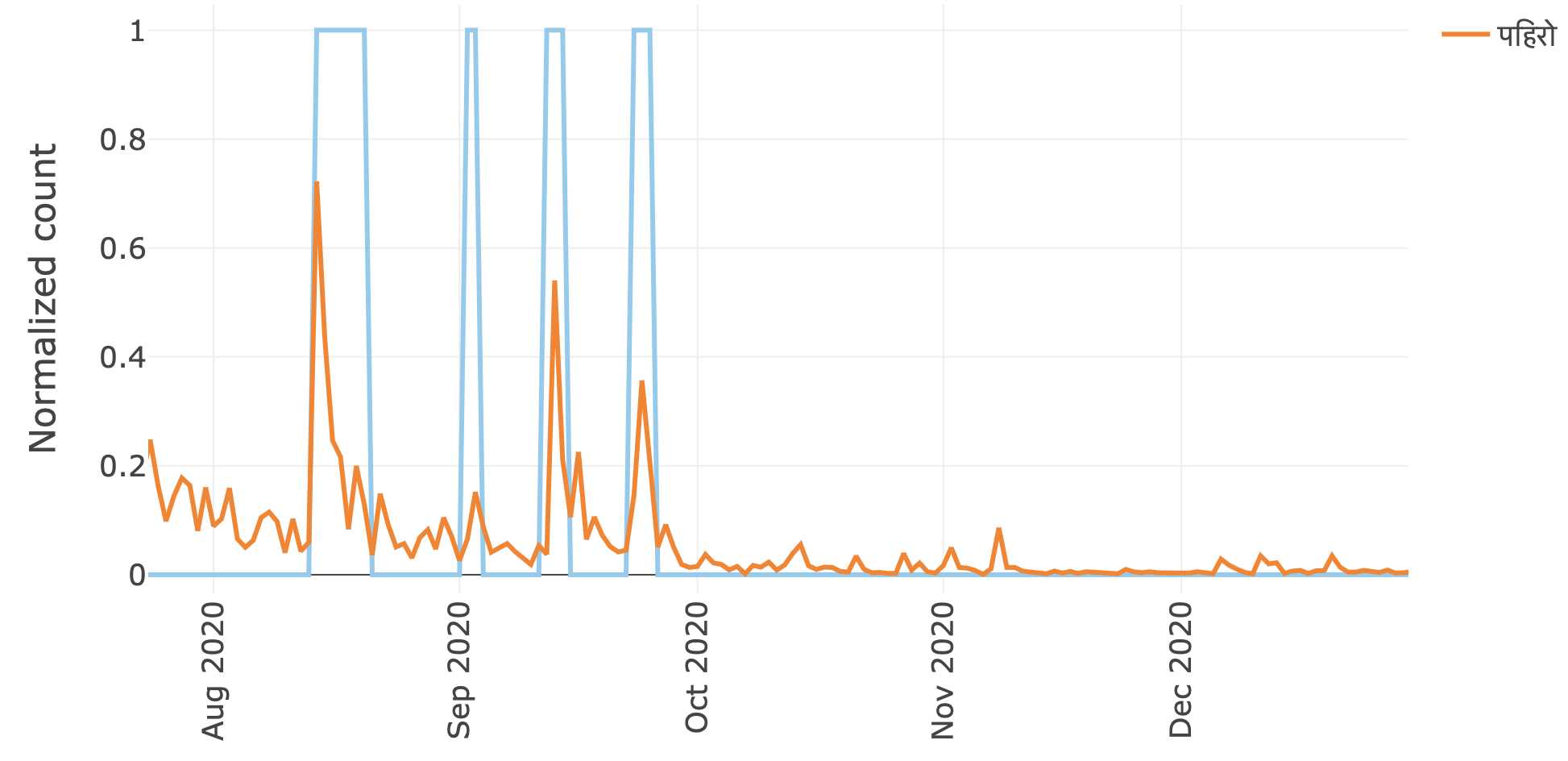}
  \caption{Occurrence of "landslide", compared with flood events in Nepal reported by GDACS}
   \label{fig:landslide}
\end{figure}

Detecting an event onset is only the first step towards awareness. Moreover, an efficient detection mechanism provides a precious byproduct: broad matching keywords related to events of interest. These, in principle, enhance the recall of the data collection. But as broad as they can be, two issues should be considered. On one side, unpredictable descriptors, event-specific qualities and linguistic variations are bound to happen. One significant example is given by location names, which are not usually constant across events. On the other side, good "general purpose" predictive features do not guarantee to match relevant contents only. In light of this consideration, adaptive methods of data collection should be prioritized. Adaptability is aimed at controlling the specificity of an event, together with its evolution over space and time. An example of an approach that is both automated and adaptive is given by \cite{Autelitano2019}, focusing on identifying variable search keywords over time. 

\subsection{Understanding and filtering semi-structured contents} \label{subsec:methods}

Methods and tools that extract structured information from social media are the main focus of this line of research. Among the general questions that can be posed, we include questions about the category of posts, as in ``\textit{is this post about} <x>\textit{?}''. These questions can be usually translated to supervised classification tasks. The questions being asked could be related to some specific semantic of the content, as in ``\textit{is there a} <y> \textit{in this image?}''. Some of these questions could still be mapped to classifications tasks, while some others relate to detection or segmentation tasks. A systematic approach for answering this kind of questions has been proposed in \cite{DBLP:conf/icse/NegriSARSSFCP21}, developing the VisualCit\footnote{http://visualcit.polimi.it:7778/} image processing toolkit, aimed at extracting indicators about emergency events. Here, image classification and object detection are usually achieved through the use of deep neural networks, both with off-the-shelf solutions and custom trained networks. In this way, specific queries can be posed to social media data, describing specific contents or situations, such as ``two or more persons'' or ``a flood event''. In addition, the ability to leverage user feedback to adjust classifiers is also a topic of interest. Pre-processing actions such as removing near-duplicates and filtering non-photographic material are also supported.

\begin{figure}[h!]
\centering
  \includegraphics[width=0.45\columnwidth]{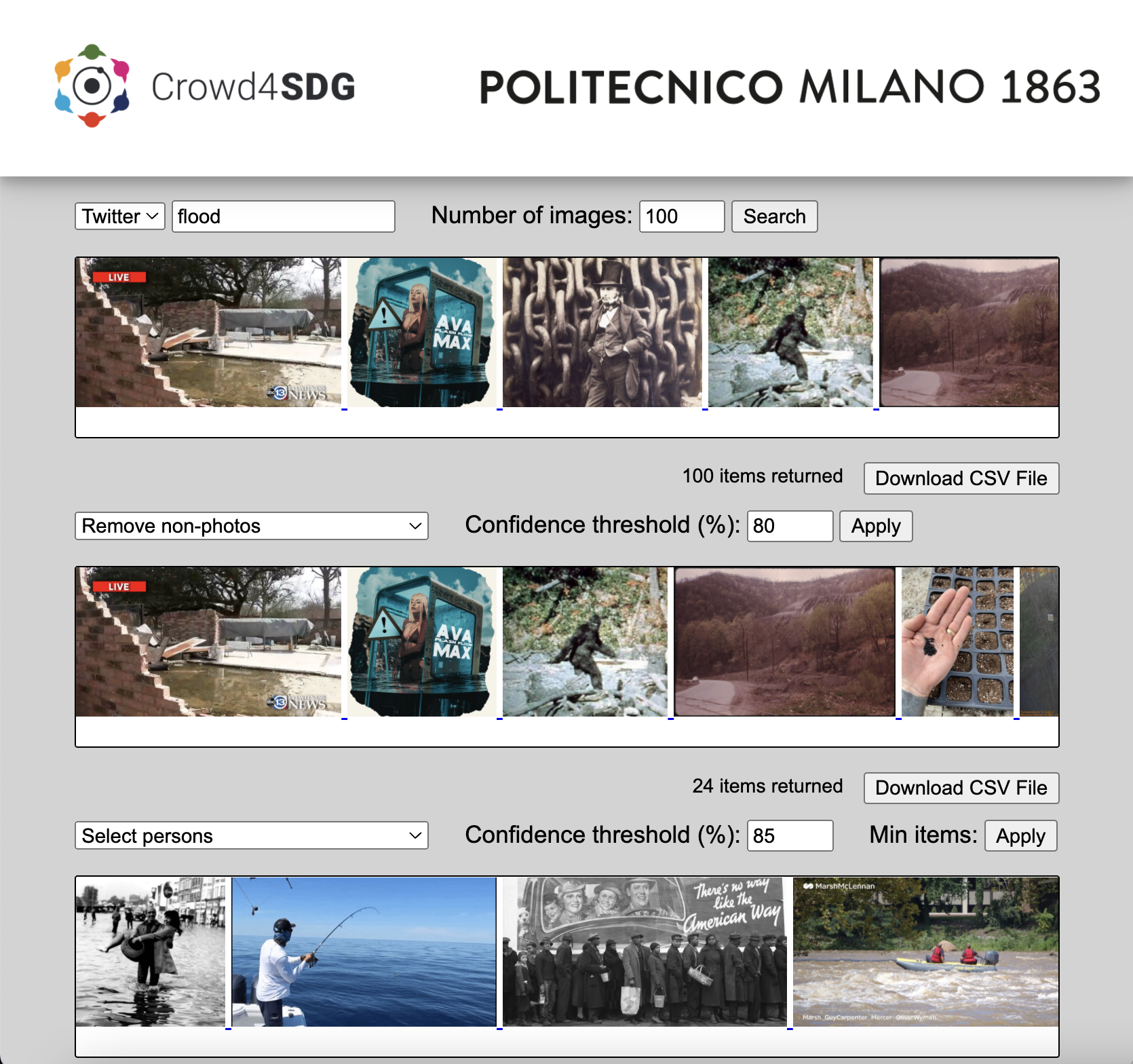}
  \caption{VisualCit interface for pipeline sketching}
   \label{fig:visualcit_demo}
\end{figure}

Text classification techniques, as well as a fused representation of textual, image, and metadata, are areas of interest, although so far they have been inspected only preliminarily. The primary goal would be to track known attributes of a situation, such as phases or topics, in order to characterize it. Since in this  case the inputs are formulated in natural language, special attention is being paid to multilingualism. Multilingualism can be supported with the use and the development of language-independent tools and representations, such as language-agnostic embeddings, or approaches that support a relevant set of languages.

Characterization of social media posts can be almost naturally attached to a temporal dimension. As a first approximation, since the extracted information augments the stream of published media, it is reasonable to assume that the information relates to near-realtime circumstances. A more pressing issue is the one about the spatial dimension of the contents. If the contents are not natively geolocated, as it usually happens\footnote{For example, for privacy issues.}, the location of the posts should be inferred. This is usually a harder problem and faceted, since the location of the author could differ from the location(s) related to the contents, or the hints for understanding the location could be flimsy. A comprehensive approach to social media geolocation has been proposed in the CIME system \cite{scalia2021cime}, which is currently being used by VisualCit. CIME disambiguates candidate locations using Nominatim\footnote{\url{https://nominatim.openstreetmap.org/}}. The disambiguation algorithm considers candidate locations, their distances, and their ranks in the OSM administrative hierarchy. CIME was exploited in the E2mC project \cite{havas2017e2mc} to support rapid mapping activities within Copernicus Emergency Management Services. Extracted geolocations can be further filtered with known geometries (e.g., countries being monitored) and/or density-based approaches, also leveraging in-post and between-post relationships.

\subsection{Building and enhancing pipelines}

\begin{figure}[t]
    \centering
    \subfigure[Is this a photo?]{\includegraphics[width=0.25\textwidth]{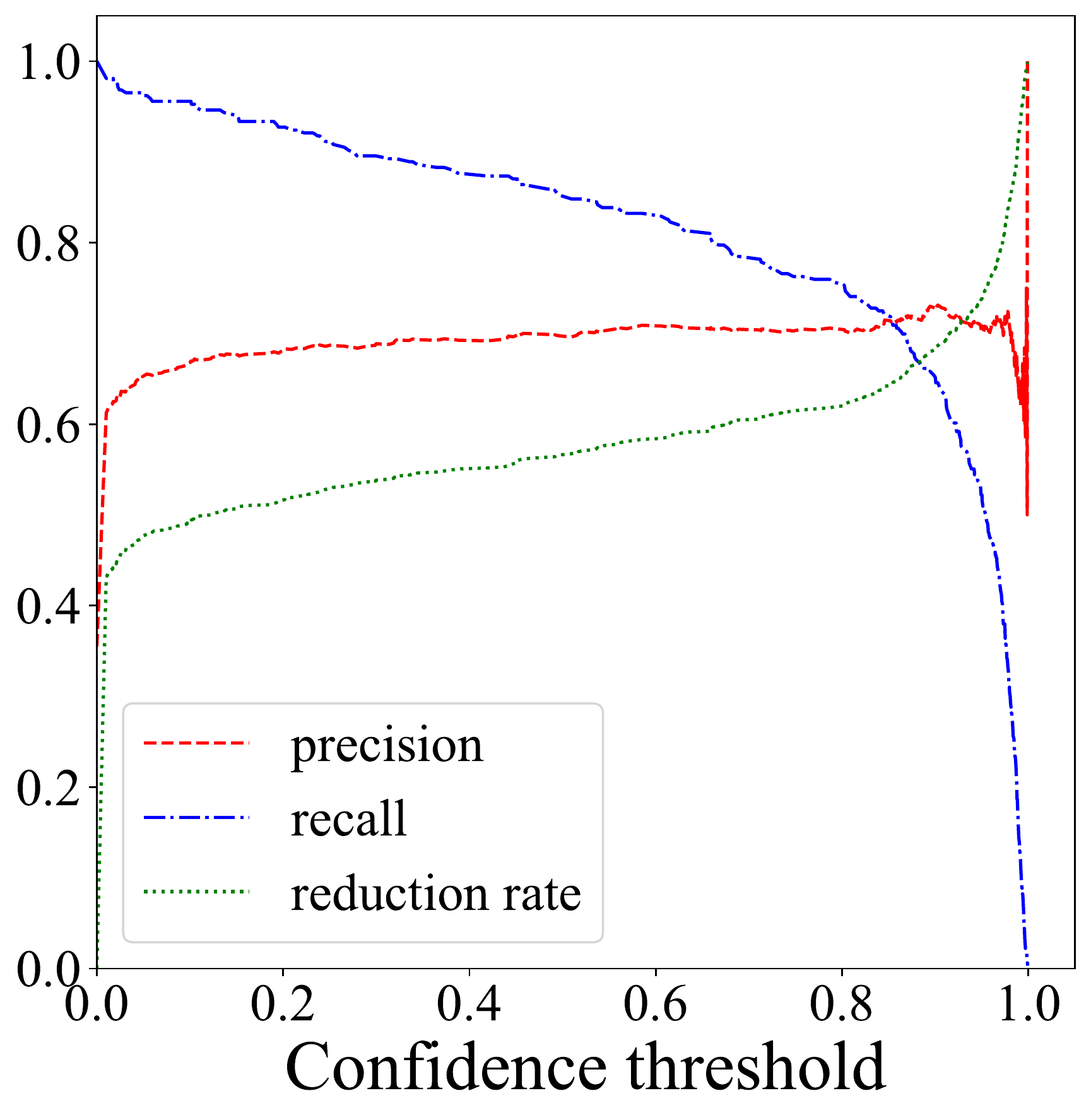}} 
    \subfigure[Are there two or more persons?]{\includegraphics[width=0.25\textwidth]{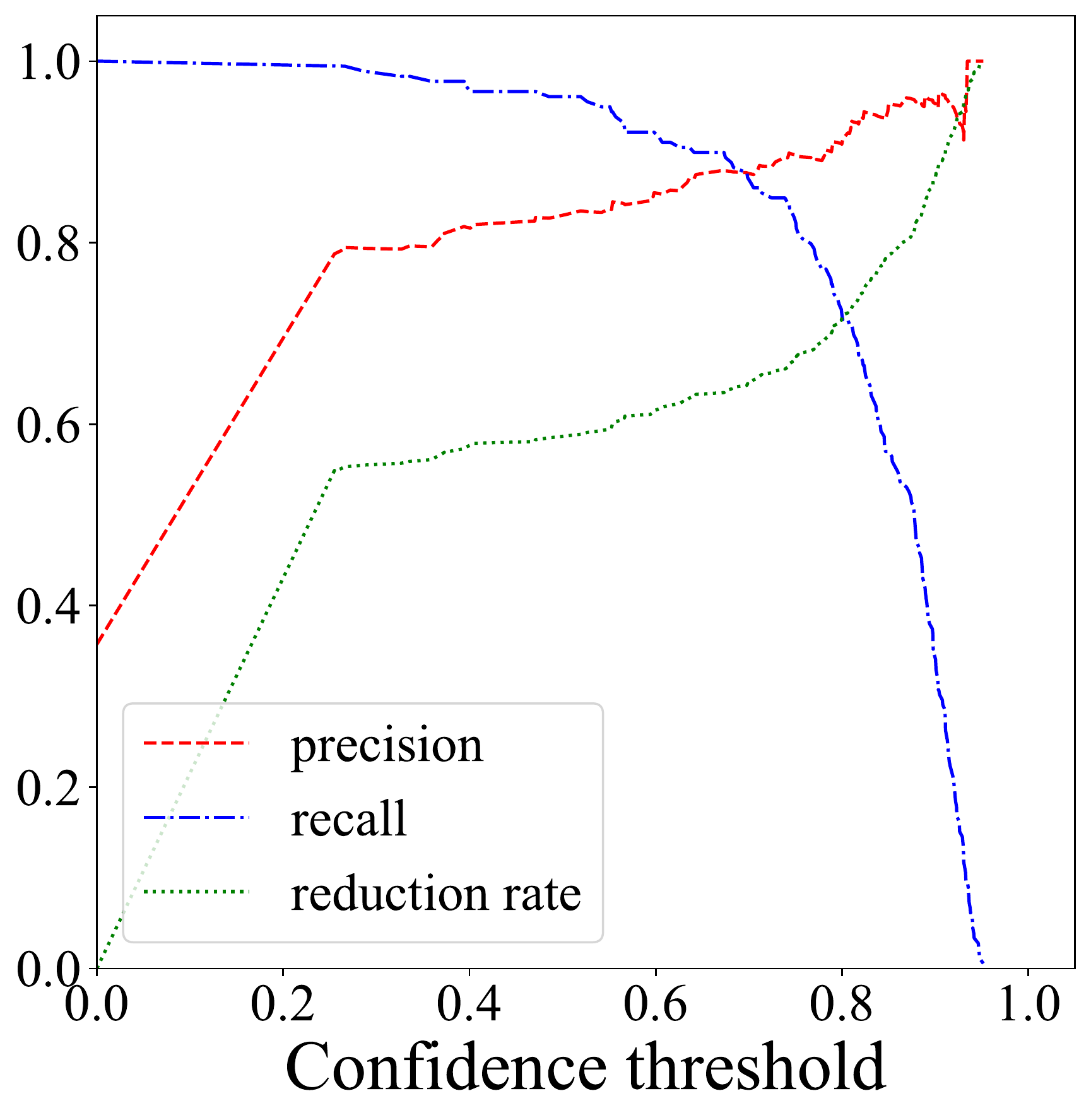}} 
    \subfigure[Is this a public place?]{\includegraphics[width=0.25\textwidth]{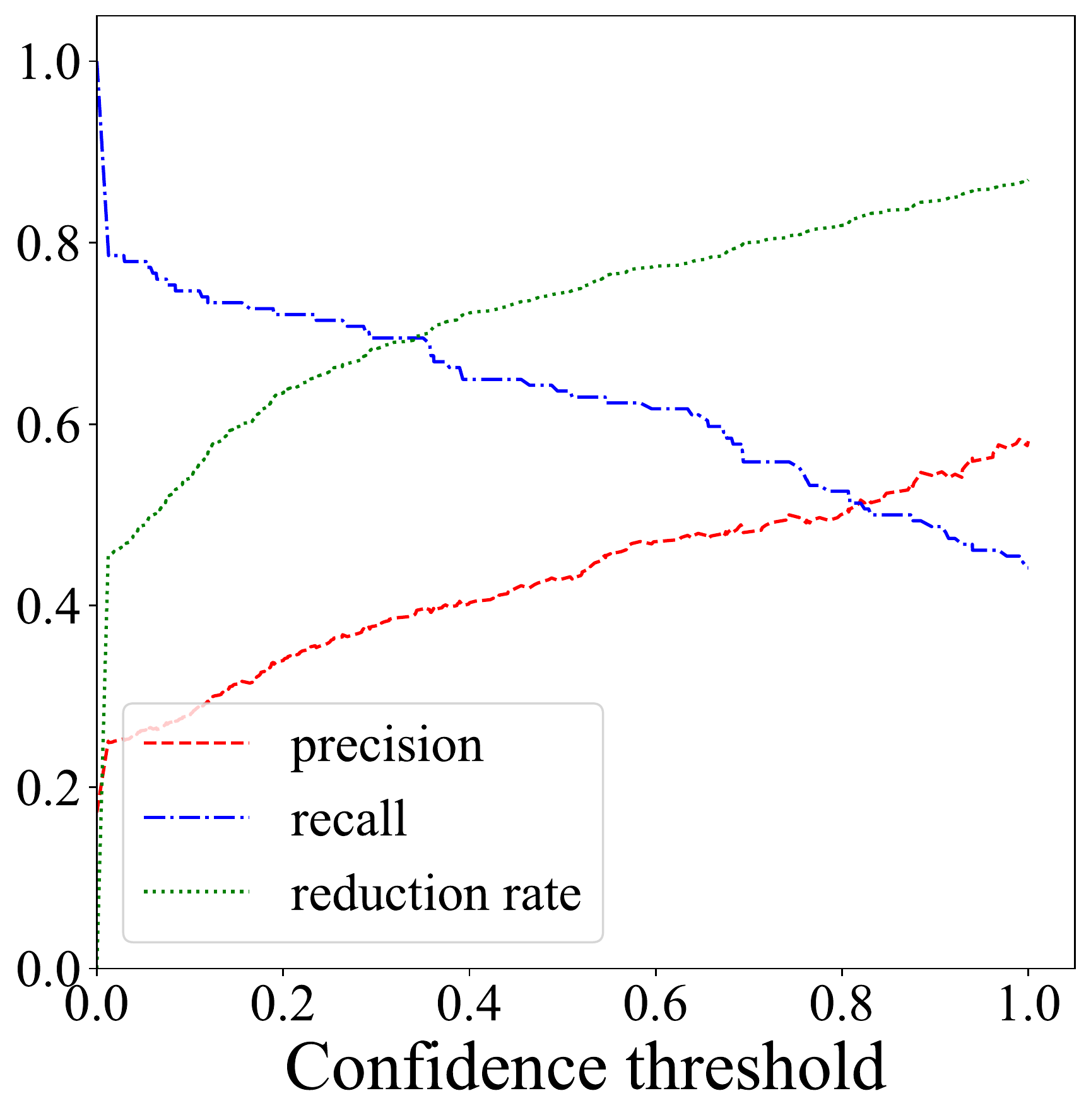}}
    \caption{Precision, recall and reduction rate responses to confidence threshold, on an example pipeline}
    \label{fig:thresholds}
\end{figure}

Comprehensive and quick reaction to large scale events requires a systematic approach to processing. Functionalities implemented so far are available with a service-based paradigm. Composite functionalities can be instantiated as a processing pipeline, using a straightforward configuration file, indicating components to be invoked and respective parameters, such as confidence thresholds. Of course, as it happens for the events themselves, the configuration of a pipeline is not always predictable. Data distributions, quality constraints and workforce availability are likely to change over time. With this fact in mind, we evaluated an iterative approach to provide suggestions to designers of data preprocessing pipelines. Iterations with user feedback are meant to rapidly achieve the desired goals, reflecting application needs and limitations and understanding the quality of the output. The designer is provided with an informative environment in which components, constraints, and parameters can be evaluated. Components typically consist of data filtering or augmentation modules, representing implementations of the tasks introduced in \ref{subsec:methods}. Constraints are usually expressed on processing quality, efficiency and cost. Figure \ref{fig:thresholds} shows an example dashboard in which the effect of the pipeline components on the final output is visualized interactively. The effect is measured in terms of precision, recall and number of filtered items. The evaluation is performed using a validated sample dataset, provided during the design process.

Processing posts with tailored pipelines dramatically reduces the number of posts. Depending on application needs, the number of items that are filtered out and the quality of the output can be balanced. Suggestions about possible improvements are provided, such as modifications to the configuration of the pipeline, based on historical or current evidence. Suggestions could also concern the execution order of the pipeline, for performance optimization. We showed that, in some test cases, an optimized configuration could lead to around 33\% of the original execution time, without affecting the output.

\section{Case studies}
\label{sec:casestudies}

As case studies, we exploited, extended and adapted available tools to assess their reliability when used as a triggering and analysis system for large scale disaster events. We applied a combination of the proposed methodologies to two flood events that occurred in 2021 in Thailand (September) and Nepal (June and July). The two events were selected since the United Nations Satellite Centre (UNOSAT) supported both activations in Nepal and Thailand with satellite-derived maps and AI-based flood detection analysis. In both  case studies, we used Twitter as a data source.

Initial evidence suggested the need for language-specific approaches rather than focusing on contents written in English. Regarding event detection, the experimentation was started with a limited amount of words. The word count signal over time proved to be sufficient to detect the two events within a reasonable time frame. Using small dictionaries is less sensitive to noise but poses a number of limitations, which have been discussed in subsection \ref{subsec:detection}. A pipeline consisting of the following tasks has been run in order to select relevant data:
  
\begin{enumerate}
    \item Remove duplicated images and similar images.
    \item Remove non-photos, such as drawings, screenshots and computer-generated images.
    \item Remove not-safe-for-work images\footnote{``\textit{Not suitable for viewing at most places of employment}'', according to the Merriam-Webster online dictionary.}.
    \item Geolocate posts using CIME.
\end{enumerate}

\begin{figure}[th!]
    \centering
    \subfigure[Removed items]{\includegraphics[width=0.35\textwidth]{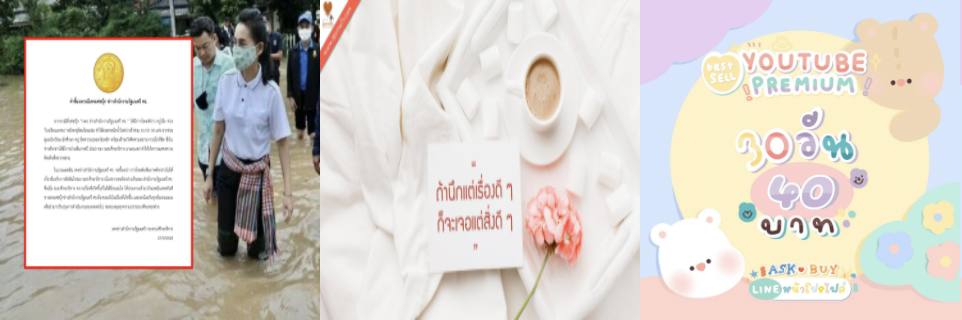}}   
    \subfigure[Kept items]{\includegraphics[width=0.35\textwidth]{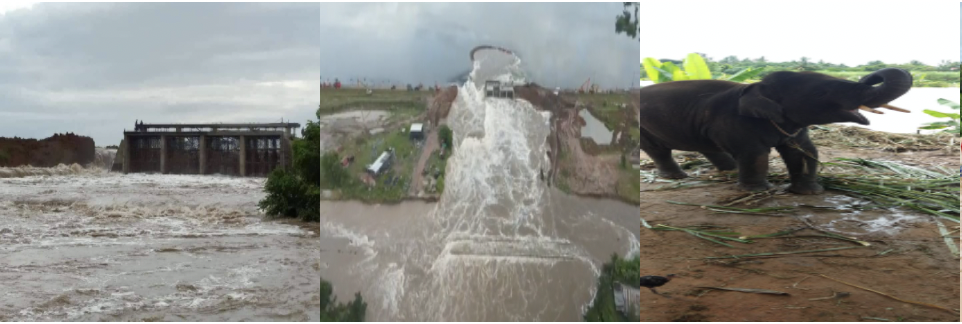}} 
    \caption{Example images for (a) items being removed by the pipeline, and (b) items being kept}
    \label{fig:thai_filtered}
\end{figure}

\begin{figure}[th!]
    \centering
    \subfigure[Geolocations / inhabitants ratio by region, normalized by population]{\includegraphics[width=0.35\textwidth]{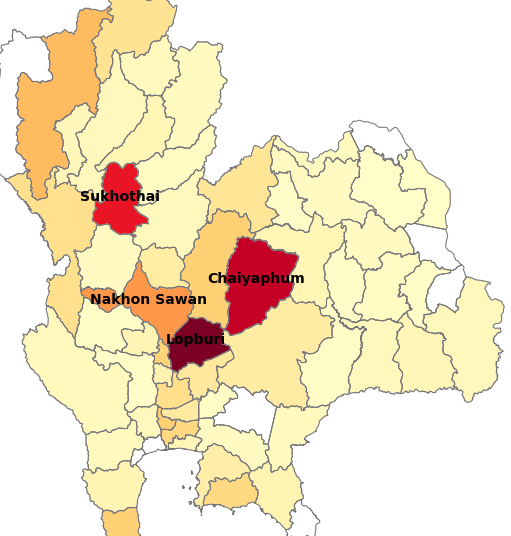}}   
    \hspace{1em}%
    \subfigure[Affected persons by region at September, 28th (source: ReliefWeb)]{\includegraphics[width=0.33\textwidth]{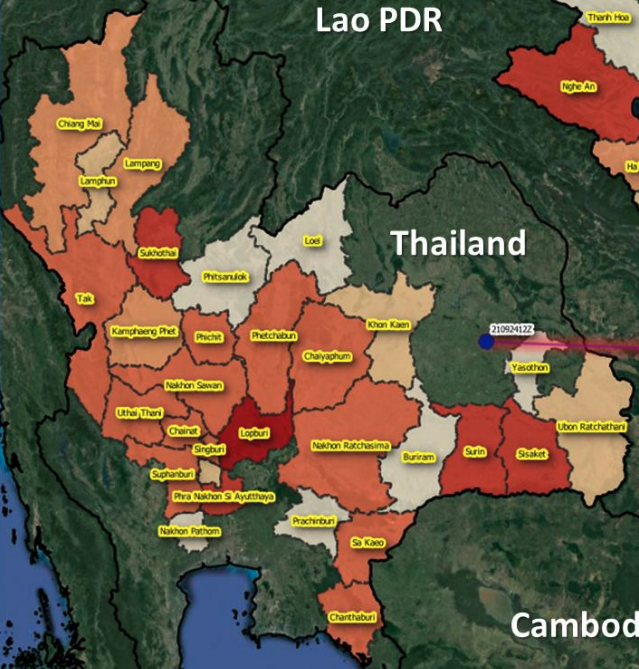}} 
    \caption{Comparing pipeline output and preliminary impact assessment, Thailand case study \cite{bono2022triggercit}}
    \label{fig:thaigeo}
\end{figure}

Processing the posts with such a pipeline reduced the amount of data by orders of magnitude. The output, aggregated by administrative region, was compared to official impact estimates (Fig. \ref{fig:thai_filtered} and \ref{fig:thaigeo}) shows promising yet improvable results. We were able to process and geolocate images at a rate of roughly 10,000 items per hour, on a single server machine. This proved to be sufficient for real-time event monitoring.

\section{Future work}
\label{sec:futurework}

The current work in progress is aimed at detection methods, components for data extraction, and process design support. Together, these elements can be leveraged to project the streams of raw data onto spatio-temporal coordinates, in order to get a suitable and actionable description of reality. So far, official, sensor and forecast data have been used as a reference for evaluating the quality of the approaches. Also, they could be fused together, in order to build a multi-faceted view of reality. This is the guiding goal for future work on the topic. At the same time, since social media contents are highly multi-modal by definition, we want to explore how different dimensions and aggregations of social media data can be combined in multi-modal representations in order to better fit the required functionalities, such as classification tasks.

\paragraph{\bf Acknowledgements}
This work was  funded by the EU H2020 project  Crowd4SDG  ``Citizen Science for Monitoring Climate Impacts and Achieving Climate Resilience'', \#872944. 

\bibliography{bibliography}

\end{document}